\documentclass[twocolumn,prd,showpacs,amsmath,amssymb]{revtex4}
\usepackage{graphicx}

\begin{document}

\title{Cosmic gamma-ray background from Type Ia supernovae revisited: 
Evidence for missing gamma-rays at MeV}

\author{Kyungjin Ahn}
\author{Eiichiro Komatsu}
\author{Peter H\"oflich}
\affiliation{
Department of Astronomy, University of Texas at Austin\\
1 University Station, C1400, Austin, TX 78712
}

\date{\today}

\begin{abstract}
The observed cosmic $\gamma$-ray background at $\sim {\rm MeV}$ has 
often been attributed to Type Ia supernovae (SNIa).
Since SNIa is close to a standard candle, one can calculate
the $\gamma$-ray intensity of SNIa integrated over redshifts fairly 
accurately, once the evolution of the SNIa rate is known. The latest SNIa 
rate measured at $z\lesssim 1.6$ (Dahlen et al. 2004) indicates that 
the previous calculations of the $\gamma$-ray background consistently 
overestimated the SNIa rate.
With the new rate, we find that the SNIa contribution is an order of
magnitude smaller than observed, and thus new population(s) of sources 
should be invoked.
\end{abstract}


\maketitle

The cosmic X-ray and $\gamma$-ray background encodes the most energetic
phenomena in the universe. It has widely been accepted that 
different population of sources contribute to the different energy bands
(see \cite{stecker/salamon:2001} for a review).
On the low energy side, the intensity spectrum in the X-ray or soft $\gamma$-ray 
region, $\lesssim 0.5$~MeV, has been successfully 
explained by the integrated counts of obscured
Active Galactic Nuclei (AGNs) \cite{comastri,zdziarski/etal:1995,ueda}. At $\gtrsim 0.5$~MeV
the spectrum of these AGNs sharply cuts off \cite{ueda}.
On the high energy side, $30~{\rm MeV}\lesssim E\lesssim 10~{\rm GeV}$, beamed 
AGNs (blazars) are able to account for
the observed background almost entirely 
\cite{salamon/stecker:1994,stecker/salamon:1996,pavlidou/fields:2002}.
The blazar spectrum appears to break below $\sim 10~{\rm MeV}$; thus, blazars
are unable to account for the low energy spectrum.
A natural question to ask is, then, ``what are the most dominant sources contributing 
to the medium energy band, $0.5~{\rm MeV}\lesssim E\lesssim 10~{\rm MeV}$?''.

Ever since the first proposal made by Clayton and Silk in 
1969\cite{clayton/silk:1969}, it has often been argued that the 
$\gamma$-ray background at
$\sim$~MeV region can be accounted for by Type Ia supernovae (SNIa)
\cite{clayton/ward:1975,the,zdziarski:1996,watanabe/etal:1999}. 
The contribution from core-collapse supernovae at $\sim $~MeV
must be much smaller than that of SNIa because the $\gamma$-ray photons
cannot easily escape the hydrogen envelope of the progenitor (a massive star)
\cite{watanabe/etal:1999}.
If this is true, the spectrum of the $\gamma$-ray background can be used 
as a powerful probe of the cosmic star formation history. 
The previous calculations of the $\gamma$-ray background from
SNIa were, however, subject to uncertainty in the supernova rate (SNR) 
of SNIa. 

In this paper, we present a more robust prediction for the cosmic
$\gamma$-ray background from SNIa using the {\it observationally determined} 
SNR, thereby avoiding any uncertainty associated with the progenitor model.
We show that the previous calculations consistently overestimated 
the SNR, and the SNIa contributes no more than 10\% of the observed level of 
the cosmic $\gamma$-ray background.
Our results strongly argue for the existence of new population(s) of sources 
contributing to the background at $\sim $~MeV.

The spectrum we derive in this paper was already presented in the previous 
paper \cite{ahn/komatsu:2005} where it is shown that the soft $\gamma$-ray 
background at $<0.511$~MeV may have a substantial contribution from
 the redshifted 0.511~MeV 
lines of dark matter annihilation in galaxies distributed over cosmological 
distances. These signals cannot, however, explain missing $\gamma$-rays
at $>0.511$~MeV.

\begin{figure}
\includegraphics[width=86mm]{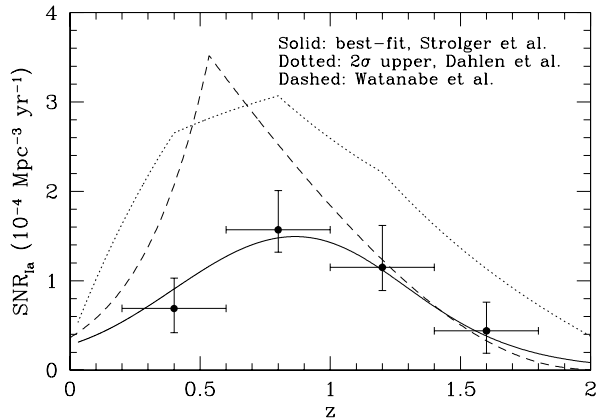}
\caption{\label{fig:snr1a}
  Type Ia supernova rates (${\rm SNR_{Ia}}$). Observed data from GOODS
  survey are shown with 1$\sigma$ statistical error bars (filled circles).
  The solid line is the best-fit model
  deduced by Strolger et al.\cite{strolger/etal:2004}. The dashed line
  is for the rates used in Watanabe et al.\cite{watanabe/etal:1999}.
  The model adopted by The et al.\cite{the} gives the SNIa rate of
  $\simeq 1.6\times 10^{-3}~{\rm Mpc^{-3}~yr^{-1}}$, which is an order
  of magnitude larger than the best-fit model and is not shown.
  The dotted
  line is constructed by connecting the 2-$\sigma$ upper limits (including
  systematic errors) of observed rates\cite{dahlen/etal:2004}. }
\end{figure}

We calculate the background intensity, $I_{\nu}$, as \cite{peacock}
\begin{equation}
 I_{\nu} =
 \frac{c}{4\pi} 
 \int 
 \frac{dz\, P_{\nu}([1+z]\nu, z)}{H(z) (1+z)^{4}},
\label{eq-inu-generic}
\end{equation}
where $\nu$ is an observed frequency, $H(z)$ is the expansion rate at
redshift $z$, and 
$P_{\nu}(\nu, z)$ is the volume emissivity
in units of energy per unit time, per unit frequency and per unit
{\it proper} volume: 
\begin{equation}
  P_{\nu}(\nu,z) = (1+z)^3~{\rm SNR_{Ia}}(z) \bar{E}_{\nu}.
\end{equation}
Here, ${\rm SNR_{Ia}}$ is the SNR of SNIa, which is
the number of SNIa per unit time and per unit
{\it comoving} volume. (Hence $(1+z)^3$ in the front.)
The time-averaged $\gamma$-ray energy spectrum of each supernova, 
$\bar{E}_\nu$, in units
of energy per unit frequency, depends 
only on $\nu$ as SNIa is a standard candle to a good approximation.
One obtains
\begin{eqnarray}
\nu I_{\nu} 
 \nonumber
 &=& \frac{c}{4\pi} \int 
 \frac{dz~{\rm SNR_{Ia}}(z)}{H(z)(1+z)^2}
 \left[ \nu(1+z)\bar{E}_{\nu(1+z)} \right]\\
 &\simeq&  
 \nonumber
 0.50~{\rm keV~cm^{-2}~s^{-1}~str^{-1}}
 \int 
 \frac{dz}{{\cal E}(z)(1+z)^2}\\
 &\times&
 \left[
 \frac{{\rm SNR_{Ia}}(z)}{10^{-4}~{\rm Mpc^{-3}~yr^{-1}}}
 \right]
 \left[\frac{\nu(1+z)\bar{E}_{\nu(1+z)}}{10^{49}~{\rm erg}}\right],
\label{eq-nuinu}
\end{eqnarray}
where ${\cal E}^2(z)\equiv \Omega_{m}h^2(1+z)^3+\Omega_\Lambda h^2$.
It follows from equation~(\ref{eq-nuinu}) that 
contribution from high redshift SNIa is negligible unless
${\rm SNR_{Ia}}(z)$ grows faster than $(1+z)^{5/2}$ toward higher redshifts.
On the other hand, recent observations suggest
that ${\rm SNR_{Ia}}(z)$ peaks at $z\sim 1$ and drops at higher redshifts; 
thus, the most dominant contribution must come from lower redshifts.
Dahlen et al.\cite{dahlen/etal:2004} have determined the ${\rm SNR_{Ia}}$
at $z\lesssim 1.6$ based on 25 SNIa observed as part
of the Great Observatories Origins Deep Survey (GOODS).
Figure~\ref{fig:snr1a} shows the data (symbols) as well as the best-fitting model
of Strolger et al.\cite{strolger/etal:2004} (the solid line).
The ${\rm SNR_{Ia}}(z)$ reaches $1.6\times 10^{-4}~{\rm Mpc^{-3}~yr^{-1}}$
at $0.6<z<1.0$ and then drops at $z>1$.
It thus follows from equation~(\ref{eq-nuinu}) that 
the expected intensity of the $\gamma$-ray background is 
$\nu I_\nu \sim 0.5~{\rm keV~cm^{-2}~s^{-1}~str^{-1}}$, as
${\cal E}(z)(1+z)\sim 2$ at $z=0.8$.

As a comparison, we also plot in Figure~\ref{fig:snr1a} the 
${\rm SNR_{Ia}}(z)$ used by the previous work\cite{watanabe/etal:1999}. 
The latest determination by \cite{dahlen/etal:2004} lies below it
essentially because their estimate was based upon
{\it indirect} determinations. 
They used an empirical analytic
function for the ${\rm SNR_{Ia}}(z)$ normalized to the local determination
of the star formation rate at $z=0$, assuming a Type II supernova rate per 
unit stellar mass of $0.007~M_\odot^{-1}$ and Type Ia to Type II ratio of 1/3. 
Another uncertainty comes from an adopted progenitor model of SNIa.
Observations suggest that ${\rm SNR_{Ia}}(z)$ follows the star 
formation rate with a delay time. 
This is where significant uncertainty in the progenitor model comes in. 
They showed that 
the uncertainty in the delay time easily affects the amplitude of the 
predicted $\gamma$-ray background by a factor of 2 to 3
(Figure~5 of \cite{watanabe/etal:1999}).
(The longer delay time reduces the amplitude.)
The ${\rm SNR_{Ia}}$ used by 
\cite{the} is even larger 
(${\rm SNR_{Ia}}\simeq 1.6\times 10^{-3}~{\rm Mpc^{-3}~yr^{-1}}$)
and is not shown in Figure~\ref{fig:snr1a}. They used an analytic model of the 
${\rm SNR_{Ia}}(z)$ for which it was assumed that all the baryons in
the universe were converted into stars. This led to a gross overestimation
of the star formation rate by an order of magnitude.
Now that ${\rm SNR_{Ia}}(z)$ has been determined up to $z\sim 1.6$, 
we can circumvent all of these theoretical uncertainties in the star
formation rate and the progenitor model, by taking observationally 
determined ${\rm SNR_{Ia}}(z)$.

Currently, the most favored scenario for SNIa 
is the explosion of a single degenerate White Dwarf close to the 
Chandrasekhar mass, in which the ignition occurs close to the center. 
Initially, the  nuclear burning front  propagates as a deflagration 
front and, after burning of about 0.2 to 0.3 $M_\odot$, turns into a 
detonation\cite{khokhlov91}. This so called ``delayed-detonation''
model allows to explain both the optical and infrared light-curves 
and spectra. This class of models produce the wide range in masses of
radioactive $^{56}{\rm Ni}$ between $0.08$  to $\approx 0.8 M_\odot$ needed 
to explain both ``normal-bright'' and subluminous SNIa, and the 
brightness decline-relation \cite{phillips93}, a cornerstone of modern 
cosmology.  For recent reviews, see  \cite{branch99,hoeflich03}.
As a typical feature of these  models (and consistent with
late time spectra \cite{Mazzali00}), the mean $^{56}{\rm Ni}$ velocities 
increase with the $^{56}{\rm Ni}$ mass and,
thus, the relative contribution of SNIa with high $^{56}{\rm Ni} $ masses 
dominate the X-ray spectra, avoiding  potential problems caused by the 
uncertainties in the  rate of subluminous SNIa
\footnote{We note that the emerging $\gamma$-ray spectrum, when averaged 
over time, is similar for all scenarios which produce bright SNIa 
\cite{hoflich/khokhlov/muller:1992}.}.
By using a SNIa at the bright-end of the distribution, we may overestimate 
the SNIa contribution to X-rays.
 
 As a reference model for the $\gamma$-ray spectrum and the light curve
of SNIa, we have chosen the delayed-detonation model, {\sl 5p0z22.23}, of 
\cite{hoeflich2002} which produces $0.561 M_\odot$ of $^{56}{\rm Ni}$ and 
represents a ``typical'' SNIa, i.e., at the bright end,
where most SNIa are observed
\footnote{Note that, with respect to the amount and distribution of 
$^{56}{\rm Ni}$,  DD-models for the ``normal-bright SNe'' are similar to the 
classical deflagration model W7 of \cite{nomoto/thielemann/yokoi:1984},
which was used in most of the previous work on the cosmic $\gamma$-ray
background \cite{the,zdziarski:1996,watanabe/etal:1999,strigari/etal:2005}.}.
The calculations are based  on a Monte 
Carlo code for $\gamma $-rays \cite{hoflich/khokhlov/muller:1992} but with  
updated bound-free opacities and nuclear branching ratios 
\cite{jundo99,berger98}.

\begin{figure}
\includegraphics[width=86mm]{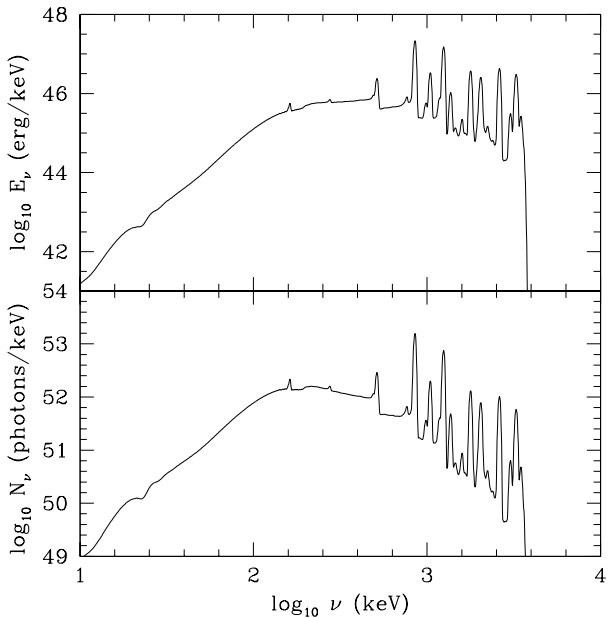}
\caption{\label{fig:spectrum}
  {\it top}: Time-averaged supernova spectrum, $E_{\nu}$, of a proto-type 
  SNIa. {\it bottom}: The same spectrum expressed in differential photon
  number (=$E_{\nu}/(h\nu)$).}
\end{figure}

Figure~\ref{fig:spectrum} shows the time-averaged $\gamma$-ray spectrum
of SNIa. The lines correspond to various $\gamma$-ray emission lines 
from radioactive decays of ${}^{56}{\rm Ni}\rightarrow {}^{56}{\rm Co}$ 
and   ${}^{56}{\rm Co}\rightarrow {}^{56}{\rm Fe}$. The line energies and
branching ratios are summarized in Table~1 and 2 of
\cite{hoflich/khokhlov/muller:1992}. All lines above 1~MeV except one
at 1.5618~MeV are due to the decay of ${}^{56}{\rm Co}$, whereas
all lines below 1~MeV except one at 0.84678~MeV are due to ${}^{56}{\rm Ni}$.

\begin{figure}
\includegraphics[width=86mm]{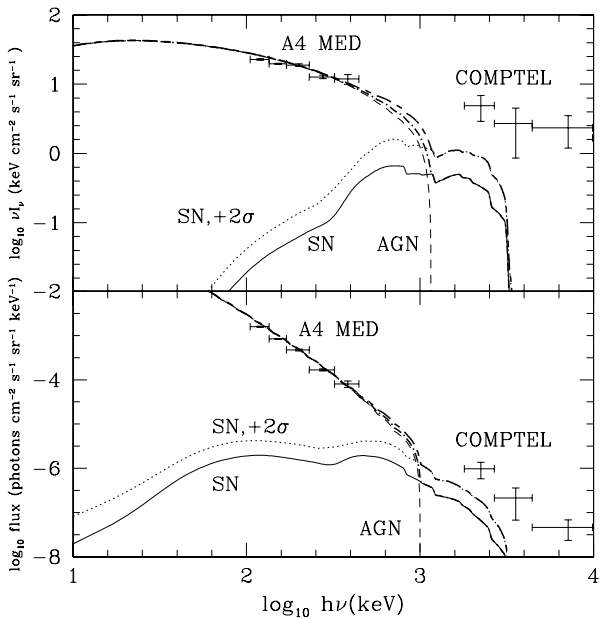}
\caption{\label{fig:cxb}
  Predicted $\gamma$-ray background from Type Ia supernovae.
  The solid line is the prediction from the best-fit ${\rm SNR_{Ia}}$ 
  by \cite{strolger/etal:2004}, while the dotted line is the one
  using the 2-$\sigma$ upper limit (including systematic errors; see also 
  Figure~\ref{fig:snr1a}). 
  The data points with error bars are from the HEAO-1 A4 MED\cite{heao} 
  and COMPTEL experiments\cite{comptel}. The AGN contribution\cite{ueda},
  which explains the HEAO data, is also plotted in the thin dashed line.
  The sum of the SNIa and AGN contributions is plotted in the thick lines.
  The top and the bottom panels use different units. }
\end{figure}

Figure~\ref{fig:cxb} shows the predicted spectrum of the cosmic
$\gamma$-ray background from SNIa as well as the observational data
from HEAO-1 A4 MED\cite{heao} and COMPTEL\cite{comptel} experiments.
The AGN contribution\cite{ueda}, which explains the HEAO-1 data at
lower energy, is also shown.
It is quite clear that the predicted SNIa signal falls short of 
the COMPTEL data by at least an order of magnitude.
We also plot the 2-$\sigma$ upper limit of ${\rm SNR_{Ia}}(z)$ 
\cite{dahlen/etal:2004} (including systematic errors),
finding that the 2-$\sigma$ limit is still a factor of 6 smaller than 
observed.
Therefore, we conclude that SNIa cannot account for the observed
$\gamma$-ray background at $\sim$~MeV, and other sources
should be invoked. 
Similar results were obtained independently 
by \cite{strigari/etal:2005}, using their ``concordance star formation
rate''.

To the best of our knowledge, there are no confirmed 
sources which could produce a substantial amount of the cosmic
$\gamma$-ray background in this energy region. 
We thus argue that there should be new population(s) of sources accounting for 
the ``missing $\gamma$-rays'' at $\sim $~MeV.
What could these sources be? 
Perhaps the most straightforward possibility would be a population
of blazars emitting in $\sim$~MeV region. These ``MeV blazars'' 
\cite{blom/etal:1995} cannot, however, be the primary candidate; otherwise one 
must require {\it all} of the regular blazars to be the MeV blazars, contrary to 
observations \cite{stecker/salamon/done:1999}.

If one had to abandon ``ordinary'' astronomical sources such as AGNs and
SNIa (which we argue we should) as an explanation to the MeV $\gamma$-ray
background, then more exotic sources would be required.
Of potential candidates would be $\gamma$-rays from dark matter annihilation, although
popular dark matter candidates (e.g., neutralinos) are usually very heavy
(dark matter mass of $m_\chi\gtrsim 30$~GeV) and it is unlikely that such heavy dark matter particles
contribute to the MeV $\gamma$-ray background \cite{elsaesser/mannheim:2004}.
Much {\it lighter} dark matter ($1~{\rm MeV}\lesssim m_\chi\lesssim 100~{\rm MeV}$),
on the other hand, is more promising, and it has recently been shown 
\cite{mev_dm} that 
such light dark matter is a promising explanation 
to the 511~keV lines detected at the center of our Galaxy \cite{spi1,spi2}.
We have shown in the previous paper that the redshifted 0.511~MeV lines from other galaxies
distributed over cosmological distances contribute to the cosmic $\gamma$-ray background
at $<0.511$~MeV substantially \cite{ahn/komatsu:2005}. Although these lines do not produce any flux
at $>0.511$~MeV, it may not be so surprising that any other associated continuum emission,
such as the internal bremsstrahlung \cite{beacom/bell/bertone:2004}, might also 
contribute substantially at $\sim$~MeV \cite{ahn/etal}. Although much is uncertain,
the MeV $\gamma$-ray background would certainly be a potential window to the new physics.

Finally, we emphasize the fact that the precise shape of the spectrum 
of the MeV $\gamma$-ray background is currently not very well constrained
(see Figure~\ref{fig:cxb}).
Given the importance of this region of the spectrum, 
it seems urgent to carry
out more precise measurements of the MeV $\gamma$-ray background.

We would like to thank D.E. Gruber for providing us with the observational
data plotted in Figure~\ref{fig:cxb}, Y. Ueda for providing
us with the AGN predictions plotted in Figure~\ref{fig:cxb},
and T. Dahlen for providing us with the 2-$\sigma$ statistical
and systematic errors  of the supernova rate (which are non-Gaussian)
plotted in Figure~\ref{fig:snr1a}. 
We would like to thank P.R. Shapiro for helpful discussion.
K.A. was partially supported by NASA Astrophysical Theory Program
grants NAG5-10825, NAG5-10826, NNG04G177G, and Texas Advanced Research
Program grant 3658-0624-1999.
The gamma-ray calculations have been performed on 11 nodes of 
the Beowulf cluster at the Department of Astronomy which 
has been financed by the John W. Cox Fund in 1999, and maintained 
by the NASA grants NAG5-7937, NASA grants HST-GO-10118.05 and 
HST-GO-10182.47 to P.A.H. 
This work was supported by NSF grant AST0307312 and LBNL 6711032 to P.A.H.


\end{document}